
%
%
\documentstyle{amsppt}
%

\def\gr#1{{\goth #1}}
%


	\def\grg{{\gr g}}

	\def\grl{{\gr l}}

	\def\grs{{\gr s}}

\def\alp{\alpha}		
\def\bet{\beta}
\def\gam{\gamma}		
\def\del{\delta}		\def\Del{\Delta}

\def\tet{\theta}

\def\ome{\omega}		
\def\nchi{\hbox{\raise 2.5pt\hbox{$\chi$}}}


\def\CalF{{\Cal F}}

\def\CalJ{{\Cal J}}

\def\CalL{{\Cal L}}

\def\CalN{{\Cal N}}
\def\CalO{{\Cal O}}
\def\CalP{{\Cal P}}

\def\CalS{{\Cal S}}
\def\CalT{{\Cal T}}


		\def\bfP{{\bold P}}


\def\gtil{\tilde g}

\def\ztil{\tilde z}

\def\Ftil{\widetilde F}


		\def\Ebar{\overline{E}}

\def\hbar{\bar h}



%
%

\def\r1{\sqrt{-1}}
\def\lb{\linebreak}

\def\lbd{\lambda}
\def\mx{\pmatrix}
\def\emx{\endpmatrix}

\def\cx{\text{\bf C}}

\def\sn{ \sum_{i=1}^n}
\def\tp{^{\text{\bf T}}}
\define\union#1#2{\overset{#2}\to{\underset{#1}\to \cup}}
\def\lo{\Cal L_{0}}
\def\li{\Cal L_{1}}
\def\lii{\Cal L_{2}}
\def\Ln{\Cal L}

\def\wt{\wtil}

\def\tr{\text{tr}}
\def\al{\alpha_i}
\def\ai{\al}

\def\kmm{\kern -6pt}
\def\ktm{\kern 7pt}
\def\kmc{\kern -7pt}
\def\ktmi{\kern 5pt}
\def\kmmi{\kern -6pt}
\def\kenmi{\kern 3pt}
\def\knenmi{\kern -4pt}

\redefine\cite#1{{\bf[#1]}}

\def\Ad{\text{\rm Ad}}

\def\contil#1{\kern2pt\bar{\kern-2pt{\tilde #1}}}

\def \smaller {\eightpoint}

\def\tr{\operatorname{tr}} 
\def\ra{\rightarrow}
\def\lra{\longrightarrow}
\def \wt{\widetilde}
\def\slr{\wt{\grs\grl}(r,\cx)}
\def\pone{\bfP^1(\cx)}
\def\Oh{\text{\bf O}}
%
%
\TagsOnRight
\parindent=8 mm
\magnification \magstep1
\hsize = 6.25 true in
\vsize = 8.5 true in
\hoffset = .2 true in
\parskip=\medskipamount
\topmatter
\title
Quasi-Periodic Solutions for \\
Matrix Nonlinear Schr\"odinger Equations${}^\dag$
\endtitle
\rightheadtext{Quasi-Periodic Solutions for Matrix CNLS Equations}
\leftheadtext{M.A.~Wisse}
\author
M.A.~Wisse${}^1$
\endauthor
\endtopmatter
\footnote""{${}^\dag$Research supported in
part by the National Sciences
Engineering Research Council of Canada and the Fonds FCAR du
Qu\'ebec}
\footnote""{${}^1$D\'epartement de math\'ematiques
et de statistique, Universit\'e de Montr\'eal,
C.P. 6128-A,\lb
Montr\'eal, P.Q., Canada H3C 3J7}
\bigskip
\centerline{\bf Abstract}
\bigskip
\baselineskip=10pt
\centerline{
\vbox{
\hsize= 5.5 truein
{\smaller The Adler-Kostant-Symes theorem yields isospectral
hamiltonian flows on the dual $\tilde\grg^{+*}$ of
a Lie subalgebra $\tilde\grg^+$ of a loop algebra
$\tilde\grg$. A general approach relating the method
of integration of Krichever, Novikov and Dubrovin to
such flows is used to obtain finite-gap
solutions of matrix Nonlinear Schr\"odinger Equations
in terms of quotients of $\tet$-functions.
}}}
\baselineskip=16pt \bigskip \bigskip
\document
\bigskip
\noindent {\bf 1. Introduction}
\medskip

In a recent paper \cite{AHH1} the method of integrating
nonlinear
partial differential equations due to Krichever, Novikov
and Dubrovin (see, for example, \cite{KN, D}) was
adapted to computing finite-gap solutions for PDE arising as
integrability conditions for Lax equations arising in  the
framework of the Adler-Kostant-Symes (AKS) theorem. Consider
a loop
algebra $\tilde\grg=\tilde\grg^+ \oplus \tilde\grg^-$, split
into the direct sum of the subalgebra $\tilde\grg^+$
of loops $X(\lbd)$, which, viewed as maps
$X:S^1 \mapsto \grg$ into a Lie algebra $\grg$ extend
holomorphically inside the unit circle
in the complex $\lbd$-plane and $\tilde\grg^-$,
the subalgebra of loops extending holomorphically
outside the unit circle, normalized by the condition
$X(\infty)=0$.
The Lax equations considered in \cite{AHH1}, such
as \thetag{1.5a,b}, determine completely integrable
Hamiltonian systems. They follow from the AKS
theorem and are are defined on the dual space
$\tilde\grg^{+*}$ of the subalgebra $\tilde\grg^+$.

In an earlier paper \cite{AHP}, flows in a finite
dimensional subspace consisting of rational elements in
$\tilde\grg^{+*}$ were related to isospectral rank-$r$
deformations of a fixed $N\times N$ matrix.
In \cite{HW} the coupled nonlinear Schr\"odinger
equations corresponding to the classical Lie algebras
(cf. \cite{FK}), which will henceforth be called
``matrix NLS equations'', were related to the \cite{AHP}
framework. The general form of these equations is
$$
\aligned
\r1 q_t&=q_{xx} + qpq \\
-\r1 p_t&=p_{xx} + pqp,
\endaligned\tag{1.1}
$$
where $q,p\tp \in \cx^{a\times b}$.

Let $\slr,\,r=a+b$ be the Lie algebra of loops in
$\grs\grl(r,\cx)$. Let $\slr^+$ be the subalgebra of
loops extending holomorphically inside the unit circle.
Using the ad-invariant inner product
$$
<X,Y>=\oint_{S^1} \frac{\tr (X(\lbd)Y(\lbd))}{\lbd}d\lbd
\qquad X,Y
\in\slr \tag{1.2}
$$
we may henceforth identify the regular part of
$\slr^*$ with $\slr$. In the
same way $\slr^{+*}$ is identified with $\slr^-_0$, the
subalgebra of loops extending holomorphically outside the
unit circle. No notational distinction will be made between
elements of $\slr^*$ (resp. $\slr^{+*}$) and
elements of $\slr$ (resp. $\slr^-_0$). Let
$\CalF_+=I(\slr^*)_{\slr^{+*}}$ be the ring of
$\Ad^*$-invariant functions on $\slr^*$ restricted to
$\slr^{+*}$. The AKS-theorem tells us that elements of
$\CalF_+$ commute in the Lie-Poison structure
of $\slr^{+*}$. Furthermore, the Hamiltonian equations for
functions in $\CalF_+$ are given by Lax pairs.
The Hamiltonians for the matrix nonlinear Schr\"odinger
equation are given by
$$
\align
\Phi_x (X)&=\frac 12 \tr\left(\left(\frac{a(\lbd)}
{\lbd^{n-1}} X(\lbd)\right)_0\right)
\tag{1.3a} \\
\Phi_t (X)&=\frac 12 \tr\left(\left(\frac{a(\lbd)}
{\lbd^{n-2}} X(\lbd)\right)_0\right)
\tag{1.3b}
\endalign
$$
for the $x$ and $t$ flows respectively, where $X\in\slr^*$
and
$a(\lbd)=\prod_{i=1}^n (\lbd-\ai)$. The $\ai$ are
constants
chosen inside the unit circle. (N.B. This restriction is
inessential; we can always choose our circle $S^1$ to have a
larger radius, so as to enclose arbitrarily chosen $\ai$'s).
Let
$$
\CalN(\lbd)=\lbd\sn\frac{\CalN_i}{\ai-\lbd} \in \slr^-_0,
\tag{1.4}
$$
where the $\CalN_i \in \grs\grl(r,\cx)$ are of fixed rank
$k_i\leq r-1$.  Hamilton's equations for \thetag{1.3a,b}
for elements of the form \thetag{1.4} are given by
$$
\align
\frac d{dx} \CalN(\lbd)=[d\Phi_x(\CalN)_+,\CalN]
\tag{1.5a} \\
\frac d{dt} \CalN(\lbd)=[d\Phi_t(\CalN)_+,\CalN]
\tag{1.5b}
\endalign
$$
where the ``$+$'' subscript means projection to $\slr^+$.
The co-adjoint orbit $\CalO_\CalN$
through $\CalN$ preserves the
pole structure of $\CalN(\lbd)$ and the rank of
$\CalN_i$ and hence these are preserved under
any Hamiltonian flow.
This orbit is therefore finite dimensional. The Lax
equations for
$$
\Ln(\lbd)\equiv\frac{a(\lbd)}\lbd
\CalN(\lbd)=\lbd^{n-1}\lo +
\lbd^{n-2} \li +\dots +\Ln_{n-1} \tag{1.6}
$$
are given by
$$
\align
\frac d{dx} \CalL(\lbd)&=[\lbd\lo + \li,\CalL] \tag{1.7a}
\\
\frac d{dt} \CalL(\lbd)&=[\lbd^2 \lo + \lbd\li +
\lii,\CalL] \tag{1.7b}
\endalign
$$
These equations may also be viewed as due to the AKS
theorem on the orbit $\CalO_\CalL$ through $\Ln$, with
Hamiltonians
$$
\align
\wt\Phi_x (X)&=\frac 12 \tr((\lbd^{2-n} X(\lbd))_0)
\tag{1.8a} \\
\wt\Phi_t (X)&=\frac 12 \tr((\lbd^{3-n} X(\lbd))_0)
\tag{1.8b}
\endalign
$$
Equations \thetag{1.1} are the integrability conditions
for \thetag{1.7a,b} if the leading terms of $\Ln(\lbd)$ are
$$
\align
\lo &= \frac {\r1}{a+b} \mx bI_a & 0 \\ 0 & -aI_b \emx
\tag{1.9a} \\
\li &= \mx 0 & q \\ p & 0 \emx \tag{1.9b} \\
\lii &= \r1 \mx qp & -q_x \\ p_x & -pq \emx \tag{1.9c}
\endalign
$$
The underlying constraints on $\lo$, $\li$ and $\lii$ are
invariant under the
flows of all Hamiltonians in $\CalF_+$.

Section 2 contains a brief summary of
the main results of \cite{AHH1}. The first part is devoted
to the properties of the invariant spectral curve arising
from Lax equations of the form
$$
\Ln_\tau = [P(\Ln(\lbd),\lbd^{-1})_+,\Ln(\lbd)] \tag{1.10}
$$
where $\Ln(\lbd)$ is a matrix polynomial of the form
\thetag{1.6}, $\tau$ is the flow parameter and
$P(z,\lbd^{-1})$ is a polynomial in $z$ and $\lbd^{-1}$.
The second part of section 2 summarizes the integration
method for equations of type \thetag{1.10} yielding solutions
in terms of $\tet$-functions.
Section 3.1 discusses the singularities of the spectral
curve underlying equations \thetag{1.7a,b}.
Solutions to \thetag{1.1} in terms of
quotients of $\tet$-functions are obtained in section 3.2.

Althogh only the generic case of $a\times b$ dimensional
rectangular matrices $q,p\tp$ is treated here, we
emphasize that finite-gap solutions for matrix NLS
equations corresponding to the various
reductions by involutive automorphisms of the hermitian
symmetric Lie algebra
$(\grs\grl(r,\cx), \grs\grl(a,\cx) \oplus \grs\grl(b,\cx)
\oplus \cx)$ (see \cite{FK, HW})
can be obtained by imposing appropriate initial conditions
on the matrix polynomial \thetag{1.6}. (See also \cite{AHH1}
for a general approach to reductions to sublagebras of
$\grs\grl(r,\cx)$).

\bigskip
\noindent {\bf 2. Preliminaries}
\smallskip

In this section we summarize the basic results of
\cite{AHH1}.\medskip

\noindent {\sl 2.1 The spectral curve for flows in a finite
dimensional rational co-adjoint orbit}
\smallskip

This subsection is devoted to the specific properties
of the spectral curve $\CalS_0$ given by
$$
\CalP(z,\lbd) = \det(\Ln(\lbd) - zI) = 0. \tag{2.1}
$$
which is invariant under flows of the Lax equation
\thetag{1.10}.

Let $\CalO(i)$ denote the i-th power of the hyperplane bundle
$\CalO(1)$ over $\pone$. The technique of \cite{AHH1}
consists in compactifying $\CalS_0$ by embedding it
into  the surface $\CalT=\CalO(n-1)$.
On $\pone$ consider standard coordinate charts $(V_0,\lbd)$
and $(V_1,\tilde\lbd)$ over $V_0=\pone\backslash\{\infty\}$,
$V_1=\pone\backslash\{0\}$, with $\tilde\lbd=\lbd^{-1}$ on
$V_0\cap V_1$. The line bundle
$\pi:\CalO(n-1)\ra \pone$ is covered by the coordinate
charts $U_i=\pi^{-1}(V_i),\,i=1,2$ with coordinates
$(\lbd,z)$ on
$U_0$ and $(\tilde\lbd,\ztil)$ on $U_1$, where
$\ztil=z\lbd^{-(n-1)}$ on
$U_0\cap U_1$. Expanding \thetag{2.1} in $z$ and $\lbd$ shows
how $\CalS_0$ may be embedded into $U_0$. Changing to
$(\tilde\lbd,\ztil)$ over $U_0\cap U_1$ extends the embedding
of $\CalS_0$ into
$U_1$ and determines a compact curve $\CalS$
on the surface $\CalT$ that coincides with $\CalS_0$ over
$U_0$. Using the adjunction formula
\cite{GH, {\rm p. 471}} the virtual genus $g$ of $\CalS$ is
easily computed to be
$$
g=\frac 12 (r-1)(r(n-1)-2). \tag{2.2}
$$
Since $\Ln$ is given by \thetag{1.6}, the curve has
singularities at the points
$(\lbd,z)=(\ai,0),\,i=1,\dots,n$ if
$k_i=$rank$(\CalN_i)\leq r-1$.
Generically we have an
ordinary $(r-k_i)$-fold intersection
over the points $\lbd=\ai$. This is the case when
the matrices $\CalN_i,\,i=1,\dots,n$ are diagonalizable.
Blowing up $\CalS$ at these points we obtain the
(possibly singular) curve $\wt\CalS$, with virtual genus
$$
\gtil = g-\sn \frac{(r-k_i)(r-k_i-1)}2. \tag{2.3}
$$
\medskip

\noindent {\sl 2.2 Periodic Solutions in terms of
$\tet$-functions for AKS-flows}
\nopagebreak\smallskip

In what follows, it is assumed that the partially
desigularized curve $\wt\CalS$ does not have other
singularities. In the case of equation \thetag{1.1}, however,
there are generically two additional singular points at
$\lbd=\infty$. Section 3 shows how solutions can be
computed considering these singular points.

Let $\CalO_\CalT (i)=\pi^* \CalO(i)$
denote the pullback of $\CalO(i)$ to $\CalT$. The
flow of a matrix polynomial \thetag{1.6}
gives rise to a flow of sheaves $\Ebar$ over $\CalT$,
defined by the exact sequence
$$
0 \lra \CalO_\CalT (-n+1)^{\oplus r}
\overset{\CalL(\lbd)-zI}\to\lra
\CalO_\CalT^{\oplus r} \lra \Ebar \lra 0.
\tag{2.4}
$$
where $\CalO_{\CalT}$ is the sheaf of holomorphic functions
on $\CalT$, identified with the trivial line bundle. By a
standard
abuse of notation we make no distinction between a line
bundle and its sheaf of sections.

If $\CalS$ is nonsingular and irreducible with genus $g$,
$\Ebar$ is a line bundle of degree $g+r-1$ over
$\CalS$. If $\CalS$ is singular and irreducible, $\Ebar$
is a pushdown by some (possibly partial)
desingularization map $\Psi:\wt\CalS \ra \CalS$ of a
line bundle of degree $\gtil+r-1$ on $\wt\CalS$.

Conversely, a linear flow of line bundles $E_\tau$ over
$\CalS$ gives rise to a flow of matricial polynomials
\thetag{1.6}, governed by a Lax equation \thetag{1.10}.

Let $\Ln(\lbd;0)$ be an initial value polynomial defining a
spectral curve $\CalS$. Let $\wt\CalS$ be the
desingularization of $\CalS$ with genus $\gtil$. Let $E_0$ be
the initial value line bundle over $\wt\CalS$ defined by
\thetag{2.4}.
The degree of $E_0$ is generically $\gtil+r-1$. If $E_\tau$
is a line bundle of degree $\gtil+r-1$ undergoing linear flow
and having $E_0$ as initial value, the tensor product
$F_\tau=E_0^* \otimes E_\tau$ is a line bundle of
degree zero and hence has a transition function $g_{01}$ from
$U_0$ to $U_1$ given by an exponential,
$$
g_{01}(z,\lbd)=\exp(\tau \mu(z,\lbd^{-1})),\tag{2.5}
$$
where $\mu$ is a polynomial in $z$ and
$\lbd^{-1}$. Fix a basis $\{\psi_\tau^1,\dots,\psi_\tau^r\}$
of sections $H^0(\wt\CalS,E_\tau)$ (viewed as functions over
a neighborhood of $\lbd=\infty$ by fixing a local
trivialization) normalized by the
condition
$$
\psi_\tau^i(\infty_j)=\delta^i_j,\ \forall \tau,\ \forall
i,j=1,\dots,r.\tag{2.6}
$$
where $\{\infty_1,\dots,\infty_r\}\in \wt\CalS$ are the $r$
points over $\lbd=\infty$.
For $p_j$ in a neighborhood of $\infty_j$ and
$\lbd=\pi(p_j)$ its projection to $\pone$, define the matrix
$\psi_\tau(\lbd)$ by
$$
(\psi_\tau(\lbd))^j_i = \psi^j_\tau(p_j). \tag{2.7}
$$
Let $z(p_j),\,j=1,\dots,r$ be the (generically distinct)
eigenvalues of $\CalL(\lbd;0)$. The matrix polynomial
$\Ln(\lbd)$ given by
$$
\Ln(\lbd;\tau) = \psi^{-1}_\tau \mx z(p_1) & & \\
& \ddots & \\ & & z(p_r) \emx \psi_\tau
\tag{2.8}
$$
then satisfies the Lax equation
$$
\frac d{d\tau} \CalL(\lbd;\tau) = [\mu(\CalL(\lbd;\tau),
\lbd^{-1})_+,\CalL(\lbd;\tau)] \tag{2.9}
$$
where the ``$+$'' subscript means projection to positive
powers of $\lbd$.

{\sl Remark.} In the AKS framework, $\mu(z,\lbd^{-1})$ is
given by $d\phi_\tau (z)$, where $\phi_\tau$ is the
Hamiltonian defining the $\tau$-flow.

To obtain a representation of the sections
$\psi^j_\tau,\,j=1,\dots,r$ in terms of $\tet$ functions,
we first describe explicitly the
sections of $E_0$. Let $\Delta$ be an effective divisor
representing $E_0$. For instance, picking a fixed vector
$v\in\cx^r$ we can take $\Delta$ to be
the sum of points
$\Delta^j=(\lbd_j,z_j)_{j=1,\dots,\gtil+r-1}$
in $\wt\CalS$ such that $(\Ln(\lbd_j)-z_jI))_{adj} v=0$, but
$(\CalL(\lbd_j)-z_jI)_{adj}\neq 0$.
Here the subscript denotes the classical adjoint
matrix and the vector $v\in \cx^r$ represents an
element of $H^0(\wt\CalS,E_0)$. Let $D_\infty=\infty_1+
\dots+\infty_r$ be the divisor consisting of the r points
over $\lbd=\infty$. Due to the Riemann-Roch theorem,
assuming $\Delta-D_\infty+\infty_i$ is in general position
there is
exactly one meromorphic function
$\psi_0^i$ on $\wt\CalS$ such that $(\psi_0^i) \geq -\Delta +
D_\infty -\infty_i$ and
$$
\psi_0^i(\infty_j)=\del_j^i.\tag{2.10}
$$
Such a function is also a section of $E_0$. The $r$ sections
$\{\psi_0^1,\dots,\psi_0^r\}$ form a basis of
$H^0(\wt\CalS,E_0)$.
Following \cite{AHH1} such sections may be expressed in terms
of quotients of $\tet$-functions defined on the Jacobi
variety $\CalJ$ of $\wt\CalS$.

Choose a basis of cycles $(a_1,\dots,a_{\gtil},
b_1,\dots,b_{\gtil})$ on $\wt\CalS$
with the intersection property $a_i\cdot a_j = 0,\,b_i\cdot
b_j = 0$ and
$a_i \cdot b_j = \del_{ij},\,i,j=1,\dots,\gtil$.
Let $\{\ome_1,\dots,\ome_{\gtil}\}$ be a basis of
holomorphic differentials on $\wt\CalS$, normalized by the
condition that
$\oint_{a_j} \ome_i = \del_j^i,\,i,j=1,\dots,\gtil$. Define
the Abel map $A: \wt\CalS\times \wt\CalS \ra \CalJ$ by
$$
A(x,y) = \left(\int_x^y \ome_1,\dots,\int_x^y \ome_{\gtil}
\right). \tag{2.11}
$$
The corresponding $\tet$ function is defined by the matrix
of $b$-periods of
$\{\ome_1,\dots,\ome_{\gtil}\}$. Define $Q^1_i +
\dots + Q^{\gtil}_i = (\psi_0^i) - D_\infty + \infty_i
+ \Del$, with
$\Del=\Del^1+\dots+\Del^{\gtil+r-1}$.
There exists $e\in\cx^{\gtil}$ such that
$$
\gathered
\tet(e) = 0 \\
\tet(e+ A(Q^j_i,y))\neq 0,\ \forall j=1,\dots,\gtil \\
\tet(e+A(\infty_j,y)) \neq 0,\ \forall j=1,\dots,r \\
\tet(e+A(\Del^j,y)) \neq 0,\ \forall j=1,\dots,\gtil+r-1.
\endgathered\tag{2.12}
$$
Fix a base point $p\in \wt\CalS$ and define constants
$$
\xalignat 2
\alp'&=\sum_{j=1}^{\gtil} A(p,\Del^j)
&\alp''&=\sum_{j=1}^{r-1} A(p,\Del^{\gtil+j}) \\
\bet&=\sum_{j=1}^r A(p,\infty_j)
&\bet_i&=A(p,\infty_i)\tag{2.13} \\
\gam_i&=\sum_{j=1}^{\gtil} A(p,Q^j_i).&&
\endxalignat
$$
Let $\del$ be the Riemann constant. With these
definitions the functions
$$
\Ftil^i(y)=\frac{\tet(A(p,y)+\del-\alp'-\alp''+\bet-\bet_i)
\prod_{j\neq i} \tet(A(\infty_j,y)+e)}
{\tet(A(p,y)+\del-\alp')
\prod_{j=1}^{r-1} \tet(A(\Del^{\gtil+j},y) + e)} \tag{2.14}
$$
define sections of $E_0$ with the correct zeros and poles.
The normalisation condition \thetag{2.10} is obtained by
setting
$$
\psi_0^i(y)=\frac{\Ftil^i(y)}{\Ftil^i(\infty_i)}.\tag{2.15}
$$

To compute sections of the time dependant line bundle
$E_\tau=E_0 \otimes F_\tau$ we recall that $F_\tau$ is given
by the exponential transition function \thetag{2.5}. Thus
sections of $E_\tau$ can be represented as functions with
zeros at $D_\infty - \infty_i$, poles at $\Del$ and
exponential singularities at $D_0=0_1 + \dots + 0_r$, the $r$
points over $\lbd=0$. Such functions are called $r$-point
Baker-Akhiezer functions.

For an explicit representaion of $\psi^i_\tau$ let
$\tilde\mu$ be a differential of
the second kind on $\wt\CalS$ with zero $a$-cycle integrals
and the same
polar part as $d\mu$. Let $U\in\cx^{\gtil}$ be given by
$$
U=\frac 1{2\pi\r1} \left(\oint_{b_1} \tilde\mu,\dots,
\oint_{b_{\gtil}} \tilde\mu \right).
\tag{2.16}
$$
Define
$$
h_\tau^i(y)=\exp\left(\tau\int_p^y \tilde\mu\right)
\frac{\tet(A(p,y)+ \tau U + \del -
\gam_i)}{\tet(A(p,y)+\del-\gam_i)} \tag{2.17}
$$
then the functions
$$
\psi_\tau^i(y) = \frac{h_\tau^i(y) \Ftil^i(y)}
{h^i_\tau(\infty_i)\Ftil^i(\infty_i)} \tag{2.18}
$$
define a basis of $H^0(\wt\CalS,E_\tau)$ satisfying
properties \thetag{2.6}.

\bigskip
\noindent {\bf 3. The Matrix Nonlinear Schr\"odinger
Equation}

\nopagebreak\smallskip
The techniques of section 2.2 also apply to the matrix
NLS equation once the corresponding spectral curve is
desingularized at $\lbd=\infty$.

\medskip
\noindent {\sl 3.1 The Spectral Curve for the Matrix NLS
Equation}

\nopagebreak\smallskip
For the matrix NLS equation the leading terms of
$\Ln(\lbd)$ are given by \thetag{1.9a-c}. In order to
study the behaviour near $\lbd=\infty$ of the
corresponding spectral curve $\CalS$, given by
\thetag{2.1} embedded into  $\CalT$, we switch to
coordinates $(\tilde\lbd,\ztil)$ which leads to the
representation
$$
\wt\CalP(\tilde\lbd,\ztil)=\det(\wt\Ln(\tilde\lbd)-
\ztil I)=0.
\tag{3.1}
$$
with $\wt\Ln(\tilde\lbd)$ given by
$$
\wt\Ln(\tilde\lbd)=\lo+\tilde\lbd\li+\tilde\lbd^2\lii+
\dots+\tilde\lbd^{n-1}\Ln_{n-1}.\tag{3.2}
$$
Expanding \thetag{3.1} in coordinates
$(\tilde\lbd,z'=-(\ztil+\r1\frac a{a+b}))$
around $(\tilde\lbd,z')=(0,0)$ gives, under invariant
genericity conditions on $\Ln_3$,
$$
\prod_{j=1}^b(z'+\tilde\lbd^3 d_j+\Oh(\tilde\lbd^4))
\prod_{i=1}^a(z'+e_i+\Oh(\tilde\lbd^2)) = 0 \tag{3.3}
$$
where $d_j,e_i$ are constants. This shows that $\CalS$
has a $b$-fold intersection of order $3$ at
$(\tilde\lbd=0,\ztil=-\r1\frac a{a+b})$. Blowing up
three times at this point reduces the virtual
genus of $\CalS$ by $\frac{3b(b-1)}2$.
Similar things
happen at the (generically) $a$-fold point of order $3$
in $(\tilde\lbd=0,\ztil=\r1 \frac b{a+b})$.
Blowing up three times at this point reduces the
(virtual) genus by $\frac{3a(a-1)}2$.

Call $\wt\CalS$ the curve obtained after desingularization
at the $n+2$ points $(\ai,0)_{i=1,\dots,n}$, $(\infty,
-\r1\frac a{a+b})$, $(\infty,\r1 \frac b{a+b})$, in
coordinates $(\lbd,z)$. Invariant genericity conditions
on $\Ln$ may be chosen such that $\wt\CalS$ does not
have other singularities. The arithmetic genus $\gtil$ of
$\wt\CalS$ is therefore
$$
\gtil=\frac 12\left((r-1)(r(n-1)-2)-\left(
\sn(r-k_i)(r-k_i-1)\right)
-3b(b-1)-3a(a-1)\right).\tag{3.4}
$$

\medskip
\noindent {\sl 3.2 Solutions for the Matrix NLS Equation
in terms of $\tet$-Functions}

\nopagebreak\smallskip
In order to use formulas \thetag{2.16-18} we need an
explicit representation of the polynomials
$\mu(z,\lbd^{-1})$ and $\nu(z,\lbd^{-1})$ defining
transition functions $g_{01}$
from $U_0$ to $U_1$ for the degree zero line bundle
$F_{x,t}$ by
$$
g_{01}(\lbd,z)=\exp(x\mu+t\nu).\tag{3.5}
$$
As mentioned before, in the AKS framework $\mu$ and $\nu$
are given by the differentials of the AKS hamiltonians
\thetag{1.8a,b}. An easy computation gives
$$
\align
\mu(z,\lbd^{-1})&=\frac z{\lbd^{n-2}} \tag{3.6a} \\
\nu(z,\lbd^{-1})&=\frac z{\lbd^{n-3}}. \tag{3.6b}
\endalign
$$

The $(x,t)$-dependance for the matrix polynomial
\thetag{3.2} is
computed as in \thetag{2.8}. The components of $\li$ carry
the solution for the matrix
nonlinear Schr\"odinger equation \thetag{1.1}. It is
recovered by differentiating \thetag{3.2}
with respect to $\tilde\lbd$ at $\tilde\lbd=0$.
The solutions $q,\,p$ to \thetag{1.1} are given by
$$
\align
q_i^j(x,t)&=\left.\frac d{d\tilde\lbd}
\right\vert_{\tilde\lbd=0}\wt\CalL(\tilde\lbd)_i^{j+a}
\tag{3.7a} \\
p_j^i(x,t)&=\left.\frac d{d\tilde\lbd}
\right\vert_{\tilde\lbd=0}
\wt\CalL(\tilde\lbd)_{j+a}^i \tag{3.7b}
\endalign
$$
$i=1,\dots,a,\,j=1,\dots,b$. A short computation
yields
$$
\align
q_i^j(x,t)&=\r1(\psi^{j+a})'(\infty_i) \tag{3.8a} \\
p_j^i(x,t)&=-\r1(\psi^i)'(\infty_{j+a}) \tag{3.8b}
\endalign
$$
where the prime designates derivation with respect to
$\tilde\lbd$. Let $\tilde\mu,\,\tilde\nu$ be normalized
differentials of the second kind on $\wt\CalS$ with same
polar part as $d\mu,\,d\nu$ respectively. Set
$$
\align
U&=\frac 1{2\pi\r1}\left(\oint_{b_1} \tilde\mu,\dots,
\oint_{b_{\gtil}} \tilde\mu \right) \tag{3.9a} \\
V&=\frac 1{2\pi\r1}\left(\oint_{b_1} \tilde\nu,\dots,
\oint_{b_{\gtil}} \tilde\nu \right). \tag{3.9b}
\endalign
$$
Fix a point $p\in\wt\CalS$ and substitute formulas
\thetag{2.14,\,2.17} into \thetag{2.18}. Use
\thetag{3.8a,b} to get
$$
\xxalignat 2
\ \ q_i^j(x,t)&=K^{j+a}_i\exp(e_i^j x + f_i^j t)
\frac{\tet(A(p,\infty_i)+xU+tV+\del-\gam_{j+a})}
{\tet(A(p,\infty_{j+a})+xU+tV+\del-\gam_{j+a})}
&&\text{ (3.10a)}\\
\ \ p_j^i(x,t)&=-K_{j+a}^i\exp(-e_i^j x  -f_i^j t)
\frac{\tet(A(p,\infty_{j+a})+xU+tV+\del-\gam_i)}
{\tet(A(p,\infty_i)+xU+tV+\del-\gam_i)}
&&\text{ (3.10b)}
\endxxalignat
$$
where the constants $K_l^k,e_i^j$ and $f_i^j$ are
given by
$$
\align
K_l^k&={\topaligned
\r1&\frac{\tet(A(p,\infty_k)+\del-\gam_k)
\tet(A(p,\infty_l)+\del-\alp'-\alp''+\bet-\bet_l)}
{\Ftil^k(\infty_k) \tet(A(p,\infty_l)+\del-\alp')}\\
&\cdot\frac{\left.\frac d{dy}\right|_{y=\infty_l}
\tet(A(\infty_l,y)+e)
\sum_{m\neq k,l} \tet(A(\infty_m,\infty_l)+e)}
{\prod_{m=1}^{r-1} \tet(A(\Del^{\gtil+m},\infty_l)+e)}
\endtopaligned}\tag{3.11a}\\
e_i^j&=\int_{\infty_{j+a}}^{\infty_i} \tilde\mu
\tag{3.11b}\\
f_i^j&=\int_{\infty_{j+a}}^{\infty_i} \tilde\nu
\tag{3.11c}
\endalign
$$
where the integrals are evaluated in the same
polygonization as that defining the Abel map $A$.

The functions \thetag{3.10a,b} are the desired
periodic solutions to the matrix nonlinear Schr\"odinger
equation \thetag{1.1}.

\bigskip
\noindent{\bf Comments}

\medskip
The solutions for the matrix NLS equations obtained above
could also have been obtained by applying the Hamiltonian
integration
method based upon computation of the Liouville-Arnold
generating function
(see \cite{AHH2,3}) determining a canonical
transformation to linearizing
coordinates. That method explicitly shows how
nonlinear Lax equations \thetag{1.7a,b} linearize on the
Jacobi variety associated to the invariant spectral curve
and makes the evaluation of many of the constants more
explicit.
However, use of the Krichever-Novikov-Dubrovin
method avoids problems associated with ``missing Darboux
coordinates'' due to the singular spectral data at
$\lbd=\infty$ implied by the sructure of the leading
terms in $\Ln$ given by \thetag{1.9a-c}.
In the Hamiltonian approach this requires the imposition
of other symplectic constraints defining a reduced
family of spectral data involving curves of the same genus
as the desingularized ones. The main shortcoming of the
present approach lies in the difficulty on explicitly
computing the constants $\gam_k$ and
$K_l^k$ in view of the complicated implicit way they arise
through formulas \thetag{2.13}, \thetag{3.9a,b} and
\thetag{3.11}.

\bigskip
\noindent{\bf Acknowledgements}

The author expresses his profound gratitude to J. Harnad,
his Ph.D. thesis advisor, for stimulating discussions and
constant support. The author also would like to thank J.
Hurtubise for providing many helpful explanations.

\bigskip
\newpage
\centerline{\smc References}

\medskip
\item{\bf [AHH1]} Adams, M.R., Harnad, J. and Hurtubise, J.,
{\sl Isospectral Hamiltonian Flows in Finite and Infinite Dimensions II.
Integration of Flows},
Commun. Math. Phys. {\bf 134} (1990), 555--585.
\item{\bf [AHH2]} Adams, M.R., Harnad, J. and Hurtubise, J.,
{\sl Liouville Generating Function for Isospectral Hamiltonian
Flow in Loop Algebras}, in  ``Integrable and
Superintergrable Systems'', ed. B. Kuperschmidt, World Scientific,
Singapore, 1990.
\item{\bf [AHH3]} Adams, M.R., Harnad, J. and Hurtubise, J.,
{\sl Darboux Coordinates and Liouville-Arnold Integration in Loop Algebras},
CRM-preprint
\item{\bf [AHP]} Adams, M.R., Harnad, J. and Previato, E.,
{\sl  Isospectral Hamiltonian Flows in Finite and Infinite Dimensions I.
Generalized Moser Systems and Moment Maps into Loop Algebras},
Commun. Math. Phys. {\bf 117} (1988), 451--500.
\item{\bf [D]} Dubrovin, B.A., {\sl Theta Functions and Nonlinear Equations},
Russ. Math. Surv. {\bf 36} (1981), 11--92.
\item{\bf [FK]} Fordy, A.P. and Kulish, P.P.,
{\sl Nonlinear Schr\"odinger Equations and Simple Lie Algebras},
Commun. Math. Phys. {\bf 89} (1983), 427--443.
\item{\bf [GH]} Griffith, P. and Harris, J., ``Principles of Algebraic
Geometry'', John Wiley, New York, 1978.
\item{\bf [HW]} Harnad, J. and Wisse, M.A., {\sl Matrix Nonlinear
Schr\"odinger Equations and Moment Maps into Loop Algebras},
to appear: J. Math. Phys. {\bf 33}(12) (1992).
\item{\bf [KN]} Krichever, I.M. and  Novikov, S.P.,  {\sl Holomorphic Bundles
over Algebraic Curves and Nonlinear Equations},
Russ. Math. Surv. {\bf 32} (1980), 53--79.

\enddocument